\documentclass[prd,aps,showpacs,amsmath,amssymb,twocolumn]{revtex4}
\usepackage{graphicx}% Include figure files
\usepackage{dcolumn}% Align table columns on decimal point
\usepackage{bm}% bold math

\begin{document}

%%%%
%    Greek Letters
%

\let\a=\alpha      \let\b=\beta       \let\c=\chi        \let\d=\delta
\let\e=\varepsilon \let\f=\varphi     \let\g=\gamma      \let\h=\eta
\let\k=\kappa      \let\l=\lambda     \let\m=\mu
\let\o=\omega      \let\r=\varrho     \let\s=\sigma
\let\t=\tau        \let\th=\vartheta  \let\y=\upsilon    \let\x=\xi
\let\z=\zeta       \let\io=\iota      \let\vp=\varpi     \let\ro=\rho
\let\ph=\phi       \let\ep=\epsilon   \let\te=\theta
\let\n=\nu
\let\D=\Delta   \let\F=\Phi    \let\G=\Gamma  \let\L=\Lambda
\let\O=\Omega   \let\P=\Pi     \let\Ps=\Psi   \let\Si=\Sigma
\let\Th=\Theta  \let\X=\Xi     \let\Y=\Upsilon

%
%%%

%%%
%    Calligraphic letters
%

\def\cA{{\cal A}}                \def\cB{{\cal B}}
\def\cC{{\cal C}}                \def\cD{{\cal D}}
\def\cE{{\cal E}}                \def\cF{{\cal F}}
\def\cG{{\cal G}}                \def\cH{{\cal H}}
\def\cI{{\cal I}}                \def\cJ{{\cal J}}
\def\cK{{\cal K}}                \def\cL{{\cal L}}
\def\cM{{\cal M}}                \def\cN{{\cal N}}
\def\cO{{\cal O}}                \def\cP{{\cal P}}
\def\cQ{{\cal Q}}                \def\cR{{\cal R}}
\def\cS{{\cal S}}                \def\cT{{\cal T}}
\def\cU{{\cal U}}                \def\cV{{\cal V}}
\def\cW{{\cal W}}                \def\cX{{\cal X}}
\def\cY{{\cal Y}}                \def\cZ{{\cal Z}}

\def\dbd{{$0\nu 2\beta\,$}}
%
%%%%

\newcommand{\Ns}{N\hspace{-4.7mm}\not\hspace{2.7mm}}
\newcommand{\qs}{q\hspace{-3.7mm}\not\hspace{3.4mm}}
\newcommand{\ps}{p\hspace{-3.3mm}\not\hspace{1.2mm}}
\newcommand{\ks}{k\hspace{-3.3mm}\not\hspace{1.2mm}}
\newcommand{\des}{\partial\hspace{-4.mm}\not\hspace{2.5mm}}
\newcommand{\desco}{D\hspace{-4mm}\not\hspace{2mm}}

%%%%

%\draft command makes pacs numbers print
%\draft

\title{\boldmath Light neutrino contribution: is it all there is to neutrinoless 
double beta decay?}

\author{Namit Mahajan
}
\email{nmahajan@prl.res.in}
\affiliation{
 Theoretical Physics Division, Physical Research Laboratory, Navrangpura, Ahmedabad
380 009, India
}

%\date{\today}

\begin{abstract}
We consider perturbative one loop QCD corrections to the light neutrino contribution to neutrinoless 
double beta decay and find
large enhancement to the rate. QCD corrections also generate structures which mimic new physics contributions
usually considered. Within some approximations, the net effect seem to almost saturate the experimental limits,
and therefore seems to implt that this is all there is to neutrinoless 
double beta decay.
\end{abstract}

\pacs{%23.40.Bw, 14.80.Fd, 14.60.St, 12.38.Bx
}
\maketitle
%\narrowtext

%\section{Introduction}
The standard model (SM) of particle physics has been extremely successful in explaining almost
all the observations, and the places where it has not been successful often have been marred with
theoretical uncertainties. However, the experimental confirmation of 
the fact that neutrinos, which are stricltly massless within SM,
are massive particles (see \cite{Tortola:2012te} for best 
fit values of the parameters) already implies physics beyond SM. Neutrinos in that sense seem to
be messengers carrying vital information about the structure of physics beyond SM. 
Neutrinos being electrically neutral opens up the possibility of them being Majorana particles
\cite{Majorana:1937vz}. This Majorana nature however can not be probed in the oscillation experiments
that have established non-zero mass of the neutrinos. An unambiguous signature of the Majorana nature
of the neutrinos would be a process, $(A,Z)\rightarrow (A,Z+2) + 2e^-$, called neutrinoless 
double beta (\dbd) decay.
Such a process violates lepton number by two units \cite{Furry:1939qr}.

On the experimental side, 
studies have been carried out on several nuclei (\cite{KlapdorKleingrothaus:2006ff}-
\cite{DeliaTosionbehalfoftheEXO:2014zza}). 
Till date however, only one of the experiments \cite{KlapdorKleingrothaus:2006ff}
(HM) has claimed observation of \dbd signal in $^{76}{\mathrm Ge}$. 
The claimed half-life is: $T^{0\nu}_{1/2}(^{76}{\mathrm Ge}) 
= 2.23^{+0.44}_{-0.31}\times 10^{25}\,
{\mathrm yr}$ at $68\%$ confidence level. By combining the Kamland-Zen and EXO-200 results, both using $^{136}{\mathrm Xe}$, 
a lower limit on the half-life $T^{0\nu}_{1/2}(^{136}{\mathrm Xe}) > 3.4 \times 10^{25}\, {\mathrm yr}$ can be obtained. 
This is in conflict with the HM claim. Recently GERDA experiment reported the lower limit 
on the half-life based 
on the first phase of the experiment: $T^{0\nu}_{1/2}(^{76}{\mathrm Ge}) > 2.1 \times 10^{25}\, {\mathrm yr}$.
A combination
of all the previous limits leads to a lower limit $T^{0\nu}_{1/2}(^{76}{\mathrm Ge}) > 3.0 \times 10^{25}\,
{\mathrm yr}$
at $90\%$ confidence level. Both the new GERDA result and the combination are again at variance with the 
positive claim of HM.
Higher statistics in future will shed more light. A different approach can be adopted where
comparison of \dbd predictions for different
nuclei can be made in order to study the sensitivity of theoretical calculations on the 
nuclear matrix elements (NMEs) used. 

Theoretically, it is practically useful to separate the \dbd decay amplitude into the long-range 
 and short-range pieces
 (for a review of theoretical and experimental issues and the sources of uncertainties and errors, 
 see \cite{Doi:1985dx} and references therein). The long range contribution arises
 when a light neutrino is exchanged while the short range part gets its name from the fact that
 the intermediate particles are all very massive and therefore the effective interaction becomes
 point-like once the heavy degrees of freedom are integrated out. The last piece of input is the 
 non-perturbative NMEs, which are the
 properly normalized matrix elements of the quark level operators sandwiched between the nucleon states.
These NMEs are the largest source
of uncertainty and the predictions can vary up to a 
factor of two or more depending upon the NMEs employed (see \cite{Simkovic:2007vu}). 

The distinction between the long range and the short range contributions to \dbd
 amplitude is also natural and appropriate from the point of view of renormalization and 
 evolution under renormalization group equations (RGEs). Till recently, the issue of RG running
 was ignored in the context of \dbd studies. It was shown in \cite{Mahajan:2013ixa} 
that perturbative QCD corrections to the short range part can have
an important effect on the \dbd rate once RG running is incorporated. This would drastically 
affect the phenomenology as well. More importantly, when different operators
are considered, some having a small strength (Wilson coefficient) at the higher scale,
and RG evolved to the relevant scale, they can become dominant, and even lead to large
cancelations. These effects can be larger than the uncertainties in NMEs that one may envisage.
Since there are a large number of models where the scale of new physics is at the TeV, which is being probed
by LHC, there is a very interesting correlation between the LHC signatures in such models and \dbd rates.
There have been many phenomenological studies in this direction (for an incomplete list see e.g. \cite{Keung:1983uu}).
In all these studies, the fact that RG evolution should be taken into consideration has been ignored. Only recently
it has been shown in the context of specific class of models that once the RG effects are
incorporated, the impact on the LHC signatures and studies can be significant \cite{Peng:2015haa}. The authors
of \cite{Gonzalez:2015ady} have considered a set of effective operators and have studied the RG evolution for that set,
and find that in some cases the limits can change by almost two orders of magnitudes. This set
is not complete as the colour mis-matched operators are not considered. A complete calculation, including
the colour mis-matched operators, will be presented elsewhere. The lesson from these studies is that the RG effects
can significantly affect the short range contributions.

The natural question then becomes: Is the long range contribution also affected by QCD corrections?
If so, how significantly?
In the present note we address this question in affirmation, as we shall see below. For the present,
we shall only concentrate on the light neutrino exchange and the SM left-handed gauge interactions.
In this context, the inverse half-life (related to the \dbd rate) can be expressed as
\begin{equation}
 [T^{0\nu}_{1/2}]^{-1} = G^{0\nu}(Q,Z)g_A^4\vert M^{0\nu}\vert^2\frac{\vert m_{\beta\beta}\vert^2}{m_e^2} \label{dbdrate}
\end{equation}
where $G^{0\nu}(Q,Z)$, $g_A$ and $M^{0\nu}$ are the phase-space factor, axial coupling and nuclear
matrix element of the process respectively. $m_e$ is the electron mass and $m_{\beta\beta} = \sum_i U_{ei}^2m_i$
is the effective Majorana mass with $U$ being the Pontecorvo-Maki-Nakagawa-Sakata (PMNS) mixing matrix and
$m_i$ being the physical neutrino masses. Uncertainties from nuclear physics reside in $G^{0\nu}(Q,Z)$
and $M^{0\nu}$, and also in the value of $g_A$ chosen. Due to the appearance of neutrino masses $m_i$ in $m_{\beta\beta}$,
\dbd decay has the potential to
 discriminate between the hierarchies of the neutrino masses. 
 The sum of the neutrino masses gets very tightly constrained from cosmology, which in turn has implications
 for the mass hierarchies. 
 
 Before discussing the structure and impact of QCD corrections, it is worthwhile to briefly outline the steps and important points
 in the calculation of the rate at the tree level ie without the QCD corrections (see \cite{Doi:1985dx}).
  \dbd amplitude has the following parts: (a) two hadronic currents and two leptonic currents - lepton part involves the 
 light Majorana neutrino propagator; (b) as mentioned above, we assume all the vertices to be the standard $V-A$;
 (c) virtual neutrino is emitted by one nucleon and absorbed by the other, with average momentum flowing through the
 neutrino propagator, set by the inter-nucleon separation, $q \sim 100$ MeV $\gg m_i \sim {\mathcal{O}}(0.1$ eV),
 implying a non-local form which decides the type of NMEs to be employed. The leptonic part  \begin{equation}
 {\mathcal{M}}_{lep}= \gamma_{\mu}(1+\gamma_5)\frac{1}{\not{q}-m_i}
\gamma_{\nu}(1-\gamma_5) \rightarrow \frac{m_i}{q^2-m_i^2} \gamma_{\mu}\gamma_{\nu}(1-\gamma_5)
 \end{equation}
 factors out and is dealt with
 separately. It turns out that (electron momenta are at most $1$ MeV $\ll q$), to a good approximation, 
 the electrons are emitted in S-wave.
  For the hadronic part, in the second order perturbation theory, one has
 \begin{equation}
 {\mathcal{M}}_{had} \propto \sum_n\frac{\langle N, final\vert J_{\mu}\vert N_n\rangle\langle N_n\vert J_{\nu}\vert N,initial\rangle}
{E_n+p_{k,elec}^0+q_{initial}^0 - E_{initial}}
 \end{equation}
 where $J_{\alpha}, J_{\beta}$ are the hadronic currents.
Since $q \sim 100$ MeV $ >> E_n-E_{initial}$ (excitation energy $\sim 10$ MeV),  one replaces
$E_n$ by an average quantity $\overline{E}$ (with this the sum over the intermediate states can be replaced with unity) - this
 is the {\it Closure Approximation} and along with the {\it Impulse Approximation} for the
hadronic currents, renders the nuclear
matrix element calculations easy.

Next consider one loop QCD corrections to the tree level \dbd diagrams. Representative diagrams are shown in Fig. 1:
box diagram and vertex
diagram.
%%%%%%%%%%%%%%%%%%%%%%%%%%%%
\begin{figure}[ht!]
\vskip 0.32cm
\hskip 1.35cm
\hbox{\hspace{0.03cm}
%\hbox{\includegraphics[scale=0.35]{fig1.pdf}}
\includegraphics[scale=0.35]{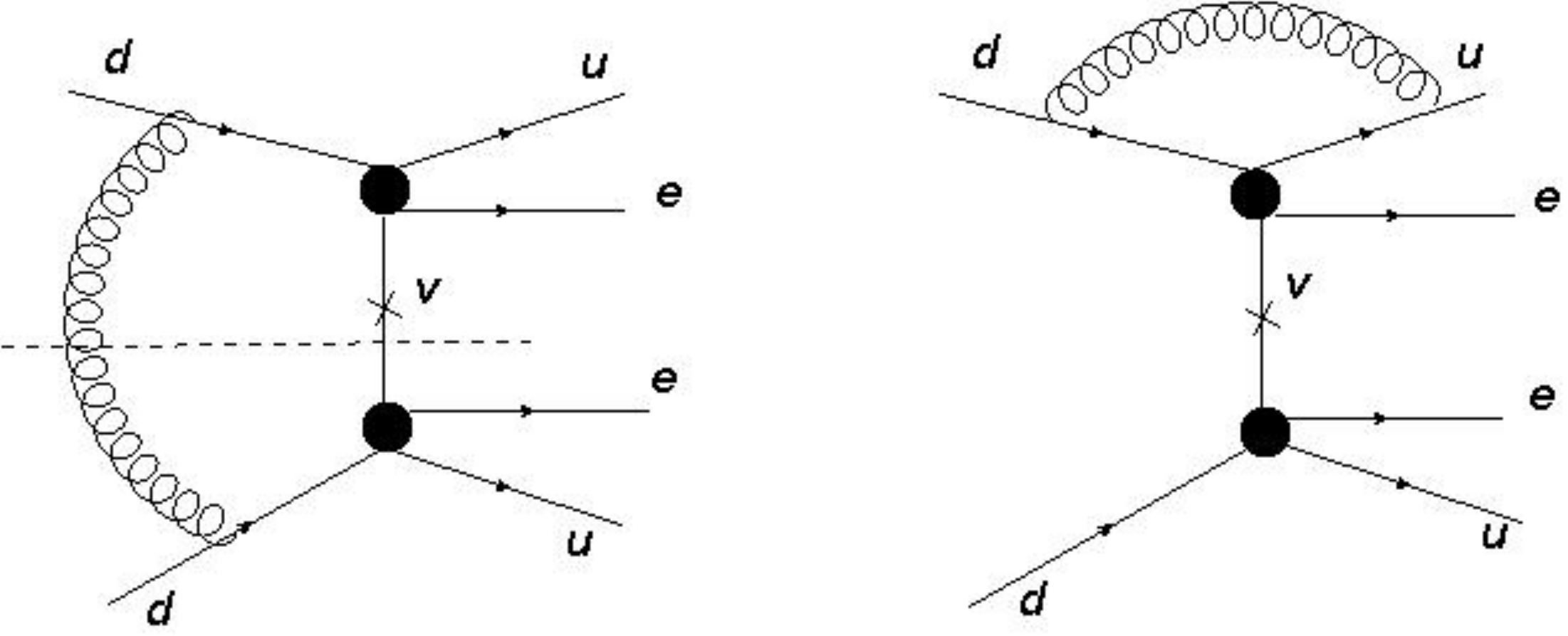}
}
\caption{Representative Feynman diagrams (drawn using the package JaxoDraw \cite{Binosi:2003yf}) showing one loop QCD corrections:
Box (Left) and vertex (Right) diagrams.
The thick circles are the standard four Fermi weak interaction vertices while the ``cross'' 
on the neutrino line denotes the Majorana nature. The horizontal dashed line denotes possible 'cut', leading to an absorptive part.
}
 \label{fig1}
\end{figure}
%%%%%%%%%%%%%%%%%%%%%%%%%%%%%%%%%%%%%
Let us consider the two type of diagrams one by one and briefly discuss the salient features of both. 
Before that, it is quite worthwhile to mention an important point that is brought out by the calculation. Due to
the loop integrals, one has to deal with complicated functions, typical of such calculations, and as a first step
one makes some approximation by neglecting some masses (like quark and electron mass) compared to a larger mass 
scale (here nucleon mass) ie one makes a series expansion in terms of the small ratio of the masses. Also,
the final nucleon momentum squared is expressed as $(1-\delta)M_n^2$, and $\delta$ is then used as an expansion
parameter after expressing all four momentum dot products in terms of $\delta$.
This is done here as well and the results presented should be looked as
first/leading term in such an expansion. This allows to have the leading terms expressed in compact analytic form.
For the box diagram, the situation is more involved and interesting. First of all, after making
the suitable expansion and retaining the leading terms, the form of the amplitude is local ie point like
rather than non-local as expected from the light neutrino exchange. The presence of a hard gluon exchanged between
the two quarks in different nucleons allows for the neutrino to go on-shell ie there is an absorptive part.
This feature is quite distinctive from the tree level amplitude where this is not possible as $q\gg m_i$.
In the intermediate stages, we use gluon mass to regulate the infra-red divergences. At the end of the calculations
this parameter is safely set to zero. Here we present the main results of the caculation. The technical details
and a systematic numerical study with specific new physics models will be presented in a separate publication \cite{namit}.\\
{\bf Vertex diagram}: it can be decomposed
into correction to tree level (${\mathcal{A}}^{(1)}$), and a new right handed current contribution (${\mathcal{A}}^{LR}$).
\begin{equation}
 {\mathcal{A}}^{(1)} \sim -\left(\frac{4 \alpha_s}{3\pi}\right)\left[\left(\frac{8}{5}+\frac{3i}{5}\right)-4i\right]
 \otimes\langle Tree\rangle
\end{equation}
where we have set the renormalization scale $\mu = M_n$ all through the calculation.
This is a large correction to the tree level amplitude.
\begin{eqnarray}
{\mathcal{A}}^{LR} &\sim& -\frac{16i\alpha_s}{3}\frac{m_i}{q^2}\left(\frac{m_um_d}{M_n^2}\right) \nonumber\\
&\otimes& \left(\frac{G_F}{\sqrt{2}}V_{ei}V_{ud}^*\right)^2\langle J_{V-A}^{had}J_{V+A}^{had}\rangle{\mathcal{M}}_{lep}
\end{eqnarray}
It is worth noting that
there was no right handed contribution to start with (only $V-A$ currents were considered) but QCD brings in a small admixture 
of right
handed contribution. This has important phenomenological implications since this QCD generated right handed 
contribution mimics the new physics
contributions typical of left-right symmetric theories. A quick comparison with expectation from left-right symmetric theory
with heavy gauge bosons $W_R\sim 3$ TeV shows that this loop induced left-right contribution is at the same level,
if not larger.\\
{\bf Box diagram}: As mentioned briefly above, box diagram generates a new structure which in the said
approximation turns out to be of the local form, though the neutrino exchanged is still a light one. In fact, the box diagram
contribution without making such an approximation does not resemble any of the forms that typically appear in \dbd calculations,
and therefore may require new NMEs which are presently not available. Within the above mentioned approximation, it reads
\begin{eqnarray}
{\mathcal{A}}^{box} &\sim& \left(\frac{2\alpha_s m_i}{M_n^2}\right)\left(\frac{2i}{3} + \frac{2}{5}\right) \nonumber \\
 &\otimes& \left(\frac{G_F}{\sqrt{2}}V_{ei}V_{ud}^*\right)^2\langle J_{V-A}^{had}J_{V-A}^{had}\rangle{\mathcal{M}}_{lep}
\end{eqnarray}
This contribution is exactly of the form of a heavy neutrino exchange but the denominator is now mass of the nucleon
rather than that of the heavy neutrino which is integrated out in the effective description to get a local
point like vertex. Or thinking in terms of light netrino, the denominator is now $M_n^2$ instead of $q^2$, but
the net effect in the end is to involve short range NMEs rather than the conventional long range NMEs for the
light neutrino exchange. This would compete with the local heavy neutrino contribution.

We can now assemble various pieces to get an idea about the impact of QCD corrections on the light neutrino
exchange contribution. First of all, the vertex corrections leads to a reasoanbly large correction to the 
amplitude - almost double. Secod, the right handed contribution is also not suppressed and the fact that
the box diagram contribution is of the local form and needs different NMEs
than the light neutrino contribution. The net effect of all this is to effectively enhance the amplitude
by a factor $2.5$-$3$ ie the total rate will get enhanced by a factor of $\sim 5$-$10$. This is a very
large correction, and part of it could be the artifact of the approximation made, namely, retaining
just the leading term in the expansion with the expansion parameter being the ratio of small masses to the 
nucleon mass. In fact, it is not possible to handle even the first sub-leading term since this forces one to
have very different nuclear physics calculations for the matrix elements. 
These are not available at the moment and may require a complete rethinking on the nuclear physics side.
The essential point is that
the structure changes due to loop integrals and one is perhaps thrown away from the closure and/or impulse
approximation. This in itself is interesting and merits a detailed investigation, beyond the scope
of the present study. Some clue may be possible if one simultaneously studies $\beta\beta$ decay with
QCD corrections included since at least the absorptive part of the box diagram can perhaps be
recast in a form close to the $\beta\beta$ amplitude by making use of the optical theorem.
One can then try to relate some matrix elements. This is quite ambitious at
this stage but worth investigating.

At this stage, let us be somewhat practical and look for the implications of such large corrections,
assuming the validity of the results obtained. Since the rate is enhanced by a large factor (called $K$),
which we take as $K=8$, like a mid-value from the given range. This is much larger
than the uncertainties in NMEs. A direct consequence of this large $K$-factor
is the decrease of the half-life $T^{0\nu}_{1/2}$ by the same amount. Recalling that different sets of NMEs
differ by a factor of $2$ or so, and the fact that any uncertainty in $g_A$ will have a large
effect as it enters as $g_A^4$ in the rate, one sees that the \dbd rate increases and therefore
the half life decreases by a factor of about $20$. This is a very large change. From the studies
on \dbd at the tree level, it is known that only quasi-degenerate neutrinos tend to saturate
the experimental limits. However, cosmological constraints on the sum of neutrino masses
already disfavour this. The inverted hierarchy predictions for the half-life are tyically less than an order of
magnitude away from the experimental limits (see eg Dev etal (arXiv:1305.0056 [hep-ph]) in \cite{Keung:1983uu}).
The impact of QCD corrections would be to disfavour a large part of the inverted hierarchy predictions as well.
For the normal hierarchy, the allowed range for the lightest neutrino mass becomes: $m_{lightest} \lesssim 0.03$ eV, while
without the QCD corrections it is about $m_{lightest} \lesssim 0.08$ eV. We therefore see that
QCD corrected rates can altogether alter all of phenomenology and the inferences we draw. 
The impact will be very significant and different on LHC related phenomenology.

Let us summarise the main points that have emerged from QCD correcting the \dbd amplitude at one loop:\\
(i) There is a large enhancement of the tree level amplitude, $\sim {\mathcal{O}}(2)$\\
(ii) Like in left-right symmetric models, there is a left-right mixed contribution which is 
generated. This contribution can be as large, if not larger than, the one in left-right symmetric
theories, and does not depend on the left-right mixing parameter and heavy gauge boson mass\\
(iii) One piece, originating from box diagrams, has a local structure and therefore requires short range
NMEs in contrast to the expected long range NMEs.\\
The net effect of all these is to drastically lower the half-life predictions - almost by an
order of magnitude. Not only is it then
possible to (almost) saturate the experimental limits simply by the light neutrino contribution,
this will have far reaching implications for phenomenology. Let us also recall that large QCD effects
have shown to be present even for the short range contributions. In the context of popular scenarios,
it was shown that \cite{Mahajan:2013ixa} there is a tendency of large cancelations between various
short range contributions. Combined with large positive corrections to the light neutrino contribution,
this on the whole appears fully consistent. Moreover, it (long range plus short range contributions
to \dbd rate) opens up the possibility that
the limits from \dbd do not come in severe conflict with LHC signatures and searches. Before closing, we would again
like to stress that the results have been obtained within some approximations, which allow a tractable
calculation at present, and it is important to try to go beyond those approximations. This however seems
to require some new nuclear physics calculations. A detailed study, including numerics, with
both the long range and short range contributions along with their QCD corrections will be presented
in a separate study \cite{namit}. It is important not to ignore the QCD corrections to \dbd, but to
incorporate them when comparing with experimental limits and in phenomenological studies.

%expt prospect
%\vskip 3cm

%MFV

%

\end{document}